\documentstyle[pictex,preprint,tighten,aps,epsfig]{revtex}
\draft
\preprint{\begin{tabular}{l}
\hbox to\hsize{July, 1999 \hfill SNUTP 99-036}\\
\hbox to\hsize{\hfill KIAS-P99060}\\
\hbox to\hsize{\hfill hep-ph/9907452}
\end{tabular} }
\begin{document}

\def\beqar{\begin{eqnarray}}
\def\eeqar{\end{eqnarray}}
\def\beq{\begin{equation}}
\def\eeq{\end{equation}}
\def\tnu{\tilde{\nu}}
\def\tS{\tilde{S}}
\def\abs#1{\left|#1\right|}
\def\ve{\langle\tilde{\nu}_e \rangle}
\def\vmu{\langle\tilde{\nu}_{\mu} \rangle}
\def\vtau{\langle\tilde{\nu}_{\tau} \rangle}
\def\vi{\langle\tilde{\nu}_i \rangle}
\def\v2{\langle\tilde{\nu}_2 \rangle}
\def\v3{\langle\tilde{\nu}_3 \rangle}
\def\vmt{\langle\tilde{\nu}_{\mu,\tau} \rangle}
\def\vS{\langle S \rangle}
\def\vN{\langle N \rangle}
\def\vev#1{\langle#1\rangle}

\title{\Large\bf Neutrino Mass and Lepton Number Violation with 
Charged Scalars}
\author{Jihn E. Kim$^{(a,b)}$\footnote{jekim@phyp.snu.ac.kr} and 
Jae Sik Lee$^{(b)}$\footnote{jslee@phys.kias.re.kr}}  
\address{$^{(a)}$Department of Physics and Center for Theoretical
Physics, Seoul National University,
Seoul 151-742, Korea, and\\
$^{(b)}$School of Physics, Korea Institute for Advanced Study, 
Cheongryangri-dong, Dongdaemun-ku, Seoul 130-012, 
Korea}
\maketitle

\begin{abstract}
If there exist only three light neutrinos below the weak scale,
the neutrino masses can arise by breaking the lepton number $L$. One 
example is the Zee mechanism for the neutrino masses. We study
phenomenological implications of neutrino masses arising from
the $L$ violation through introduction of $SU(2)\times
U(1)$ singlet scalars with $Q_{em}=1$ and 2.
\end{abstract}


\def\a{${\ddot {\rm a}}$}

\newpage

\section{Introduction}

Historically, neutrinos contributed significantly in the development
of particle theory. Firstly, Pauli postulated it
to satisfy the energy conservation and rotational invariance.
Then weak interaction with neutrinos was described by Fermi's
four fermion interaction which was known to be nonrenormalizable. Thus the 
current renormalizable theory of weak interactions, standard model(SM), 
can be traced to the origin where the introduction of the electron 
type neutrino, $\nu_e$, was crucial. However, neutrinos
have been elusive, which made it difficult to find their properties
such as the masses and magnetic moments.   

We are now in the new era of neutrinos with the accumulating
evidence on their oscillation~\cite{solar,atm1,atm2,LSND}. In particular,
we have some information on $\Delta m^2_{ij}$ and mixing angles.
These important data, especially the maximal mixing of the muon type
neutrino $\nu_\mu$ with a neutrino other than $\nu_e$,
\begin{equation}
{\rm a\ mass\ eigenstate}\simeq\frac{1}{\sqrt{2}}\nu_\mu
+e^{i\alpha}\frac{1}{\sqrt{2}}\sum_{i} c_i\nu_i,\ \ (i\ne \nu_e,\nu_\mu)
\end{equation} 
may hint a new particle(s) or a new theory.

Because it has been turned out that the mixing angles are almost
maximal, it is worthwhile to study the neutrino mass 
generating mechanisms without using the see-saw mechanism.
To pose the problem within the zoo of the known SM fermions, we
assume that for weakly interacting fermions there exist only 
the SM fermions below the electroweak scale. But we will 
introduce scalars if needed. [With supersymmetry,
the fermions, axino~\cite{axino} and 
gravitino~\cite{gravitino}, can be light since their interactions
are much weaker than the weak interaction.] Namely we are assuming
three light neutrinos: $\nu_e,\nu_\mu$ and $\nu_\tau$. We do not
assume light sterile neutrinos. Then there does not exist a Dirac
mass term for neutrinos. Neutrino masses must be of the
Majorana type,
\begin{equation}
-m_{ij}\nu_i^T\nu_j,\ \ (i,j=e,\mu,\tau)
\end{equation}
where the matrix $C^{-1}$ is omitted for the simplicity of notation.
The Majorana neutrino mass term violates the lepton number $L$. Thus
the neutrino masses can arise only if the theory violates the lepton 
number. The well known see-saw mechanism\cite{seesaw} violates $L$ 
through the Majorana mass term at high energy scale which is the
source for the light neutrino masses. 

In this paper, we study phenomenological implications of the
neutrino mass of a model which does not have a true Goldstone boson.
Toward a model with this property, we will not introduce a
neutral scalar field. Then in the vacuum respecting $U(1)_{\rm em}$
invariance, there is no place for a Goldstone boson. 
Of course, with neutral scalars introduced, 
the potential can have appropriate
parameters so that the additional hypothetical neutral scalars do not 
develop VEV's. We simply do not bother to worry
about to find this limited region of the parameter space.

In Sec.~II, we introduce a few models in which neutrino masses
have been generated. In this short review, we emphasize the symmetry
argument and relate it to Feynman diagrams. In Sec.~III, we discuss
a model without neutral Higgs scalar field to avoid possible problems
with Goldstone bosons. In this model, we present the limits
of the parameters introduced from various experimental data.
Sec.~IV is a conclusion. 
  
\section{A Short Review on Neutrino Mass with Scalars}

For the neutrino mass, it is important to pinpoint how
the $L$ symmetry is broken in the model. In the
literature, already there exist numerous studies on
the $L$ symmetry violation through singly charged scalars.
In this section, we briefly comment on the diagrammatic
symmetry argument in these models.

\subsection{Triplet Higgs Scalar}

Majorana neutrino mass can be generated with a triplet Higgs scalar 
$\xi=(\xi^{++},\xi^+,\xi^0)$ through the interactions \cite{GR}
\beq
f_{ij}\left[\xi^0 \nu_i \nu_j +\xi^+ (\nu_i l_j +l_i \nu_j)/\sqrt{2}
+\xi^{++} l_i l_j+{\rm h.c.} \right],
\eeq
where $f_{ij}$ is symmetric under the exchange of indices. If we assume
$L$ symmetry, $L(\xi)=-2$.
To generate neutrino mass $L$ must be broken spontaneously, i.e.
$\langle \xi^0 \rangle \neq 0$. 
We take the following potential for the triplet and
one Higgs doublet $H=(H^+,H^0)$
\beq
V = m^2 H^{\dagger} H + M^2 \xi^\dagger \xi +\frac{1}{2} \lambda_1
(H^\dagger H)^2 + \frac{1}{2} \lambda_2 (\xi^\dagger \xi)^2
+\lambda_3 (H^\dagger H) (\xi^\dagger \xi).
\eeq
where we have not allowed the cubic term,
\beq
\mu(\bar{\xi}^0 H^0 H^0+\sqrt{2} \xi^- H^+ H^0 + \xi^{--}H^+ H^+)+{\rm h.c.}.
\eeq 
With the potential Eq. (4), the vacuum expectation values (VEV) of
$\langle H^0 \rangle$ and $\langle \xi^0 \rangle$ can be developed.
The presence of non--zero VEV of the neutral triplet Higgs boson modifies the
$\rho$ parameter as follows \cite{GR,erlan}:
\beq
\rho \equiv \frac{M_W^2}{M_Z^2\cos^2\theta_W} =
\frac{1+2\langle \xi^0 \rangle^2/\langle H^0 \rangle^2}
{1+4\langle \xi^0 \rangle^2/\langle H^0 \rangle^2} \, .
\eeq
{}From $\rho_{\rm exp}=1.00412 \pm 0.00124$ \cite{peccei,rhoexp}, 
one can obtain
$\langle \xi^0 \rangle \leq 4$ GeV. 
To introduce an eV range neutrino mass, $\langle\xi\rangle\sim$~O(eV).
This means that the parameter $M^2$ must be
fine--tuned to equal
to $(\lambda_3/\lambda_1) m^2\sim\ {\rm eV}^2$~\cite{mautpal}. 
This is unnatural. Moreover,
this model contains the massless CP--odd field (majoron) which corresponds
to the Goldstone boson due to the spontaneous breaking of $L$ and CP--even
field which has a small
mass proportional to $\langle \xi^0 \rangle$. This leads to the
$Z$ decay into the majoron and the light scalar with a decay width of two
neutrino flavors \cite{ggn}, which is ruled out because of the LEP results
of $N_\nu=2.994\pm 0.011$ and $\Gamma_{\rm inv}=(500.1\pm 1.9)$ MeV
\cite{peccei,rhoexp}. But this model with the cubic terms given in Eq.~(5),
which violates $L$ explicitly, can be considered natural since the
vacuum expectation value is of order
$\langle \xi^0 \rangle \sim \mu \langle H^0 \rangle^2 /M^2$
in the limit $\langle H^0\rangle/M\ll 1$
and consistent with the LEP data because all physical neutral scalars become
heavy \cite{mautpal,cubictriplet}.
 
\subsection{The Zee Model}

Another $L$ violating model was proposed by Zee~\cite{zee},
\begin{equation}
{\cal L}={\cal L}_{\rm SM}+\epsilon_{\alpha\beta}l^T_{\alpha i}l_{\beta
j}\phi^+ + \epsilon_{\alpha\beta}H_{1\alpha}H_{2\beta} \phi^+ 
+{\rm h.c.}
\end{equation}
where $l_i$ is the $i^{\rm th}$ lepton doublet, $H_1$ 
with $Y=-1/2$ is the Higgs doublet
present in the SM Lagrangian ${\cal L}_{\rm SM}$, $H_2$ with
$Y=-1/2$ is the Higgs doublet {\it not coupled to fermions}, $\phi^+$
is a singly charged $SU(2)$ singlet scalar field, $\epsilon_{\alpha
\beta}$ is the $SU(2)$ Levi-Civita symbol, and the couplings are
suppressed. The coupling $l_il_j\phi^+$ is antisymmetric in $(i,j)$.
The SM Lagrangian gives $H_1$ the lepton number $L=0$. 
The $ll\phi$ coupling defines
$L=-2$ for $\phi^+$. The $H_1H_2\phi^+$ coupling defines $L=+2$
for $H_2$. Thus the above Lagrangian does not violate
the $L$ number. But $L$ can be broken by the vacuum expectation 
value(s) of the $L$ carrying neutral Higgs field(s). Indeed,
there exist one such component in this theory, the neutral
element of $H_2$. Thus, $\langle H_2^0\rangle\equiv v_{LV}\ne 0$
breaks $L$ spontaneously, and this theory predicts a Goldstone
boson. Neutrinos obtain mass at one loop-level, as shown in
Fig.~1. Note that Fig.~1 includes all the couplings and the VEV
needed to break $L$ as discussed above.
To give the Goldstone boson a mass, the Lagrangian should
violate the lepton number explicitly. It can be done by introducing 
$m_{12}^2\tilde H_1H_2+{\rm h.c.}$ where $\tilde H_1=i\tau_2 H_1^\dagger$.
Then the neutrinos get masses through
the two-loop diagrams as in Fig. 2 even with $\vev{H_2^0}=0$.
In any case, the diagonal elements of the mass matrix vanish.

The phenomenological consequences of the model has
been studied extensively in the literature~\cite{zee1,smitani}. 
One simple feature
is that the neutrino mass matrix has vanishing diagonal elements
in the flavor basis, which can lead to a
nontrivial prediction on the neutrino oscillation phenomenology.  


\subsection{Supersymmetry with R-parity Violation}

In this section, we briefly comment on the roles of scalars(mainly
the sleptons) in the R-parity violating models \cite{rpv1,rpv2}. 
But for the
baryon number conservation, we impose
the baryon parity ($B$-parity).
Then there exist the $L$ violating terms in the superpotential
in the minimal supersymmetric standard model (MSSM),
\beq \label{lvsup}
\mu_i \hat{H}_2 \hat{L}_i + 
{1\over2}\lambda_{ijk} \hat{L}_i \hat{L}_j \hat{E}^c_k +
\lambda'_{ijk} \hat{L}_i \hat{Q}_j \hat{D}^c_k \,.
\eeq
These interactions lead to the neutrino mass through the tree and one-loop
level as shown in Fig.~3 with the appropriate gaugino 
mass term $M_{\tilde Z} {\tilde Z} {\tilde Z}$ and the
soft terms $A^l {\tilde e} H_1^0 {\tilde e^c}$ and
$A^d {\tilde d} H_1^0 {\tilde d}^c$.

In the tree level diagram, the gaugino mass term and the gaugino
interaction term with the neutrino and sneutrino are needed 
to define the appropriate lepton number of the sneutrino field.
The gaugino mass mass term defines $L({\tilde Z})=0$; 
and hence $L({\tilde \nu})=1$. Thus $\Delta L=2$ results from two
insertions of $\langle\tilde\nu\rangle$. Fig.~3(a) contains all
these information. Being the gauge interaction, the
$\tilde Z$ coupling is universal (democratic), and the mass matrix
arising from this diagram contains the common entry to every elements.
Thus, only one state obtain mass by Fig.~3(a).

Similarly, the loop diagrams present in Fig.~3(b)(c) violate
the lepton number and contribute to the neutrino masses.
At one-loop level, $A^{(l,d)}$ term and the mass term of the charged lepton
(down-type quark) are needed to define the lepton numbers of the $H_1^0$,
${\tilde e}$, and ${\tilde e^c}$ fields. In the case of $\lambda_{ijk}$ as shown
in Fig.~3 (b), the charged lepton mass term and $A$ term give
$L(H_1^0)=0$ and $L({\tilde e})+L({\tilde e^c})=0$, respectively. 
This lepton number
assignments leads to $|\Delta L|=2$ interactions through
$\nu e {\tilde e^c}$ and $\nu e^c {\tilde e}$ interactions. This explains
that why, at least, four kinds of interactions are involved to generate the
Majorana neutrino mass and the induced mass is proportional to the internal
fermion mass and $A$ term. Similar arguments can be applied for the case of
$\lambda'_{ijk}$ case, viz. Fig.~3 (c).
Since the couplings appearing in these diagrams are the Yukawa
couplings, the mass matrix arising from the one-loop is
quite general and all three neutrinos can obtain masses.

\section{A Model with $\phi^+$ and $\Phi^{++}$}

In the remainder of this paper, we study another interesting model, 
violating $L$ with scalars, singly charged
$\phi_a^+\ (a=1,2)$ and doubly charged $\Phi^{++}$ \cite{babu}.
This model is free from the problem of a Goldstone boson
since there is no extra neutral scalar field.
We can generate the neutrino masses with just one singly
charged scalar $\phi_1^+$ and 
one doubly charged scalar $\Phi^{++}$. Then
the resulting neutrino mass matrix is not general enough, and hence we
introduce an additional singly charged scalar  
\begin{equation}
{\cal L}={\cal L}_{SM}+ f^a_{ij}\epsilon_{\alpha\beta}l^T_{\alpha i}
l_{\beta j}\phi_a^++\mu_{ab} \phi_a^+\phi_b^+\Phi^{--}+
\lambda^{ij}_{--}e^{cT}_ie^c_j\Phi^{--}+{\rm h.c.}
\end{equation}
where $a,b=1$ or 2, $i$ is the family number index $i=1,2,3$, 
and $e^c$ is the $SU(2)$ singlet charged anti-lepton
field. In a theory with only one singly charged $SU(2)$
singlet scalar, we have $a=1$ only. Note that we introduced an
$SU(2)$ singlet doubly charged scalar $\Phi^{++}$.
Note that coupling matrix $f^a$ is an antisymmetric matrix and $\mu_{ab}$ is
symmetric. 
The $f$ coupling
defines the $L$ number for $\phi_a^+$ as $-2$. The $\mu$ coupling
defines the $L$ number for $\Phi^{++}$ as $-4$. But the $\lambda$
coupling defines the $L$ number of $\Phi^{++}$ as $-2$. Thus, the
interaction terms give inconsistent $L$ numbers of $\Phi^{++}$, i.e. the
Lagrangian does not respect the $L$ symmetry. Since $L$ is
not a symmetry of $\cal L$, neutrino masses can arise at higher
orders, here at a two-loop level. At least the two-loop as 
shown in Fig. 4 is needed to include $f,\mu$ and $\lambda$ couplings, 
which are the requisite for the violation of $L$.
Because $L$ is explicitly broken in the Lagrangian,
there does not exist a Goldstone boson in this model.

{}From the above Lagrangian and Fig. 4, 
we estimate the two-loop neutrino mass as
\beq
\left(m_\nu^{\rm 2-loop}\right)_{im} \approx
\sum_{a,b,j,k}\lambda_{--}^{jk}f^a_{ij}f^b_{mk}
\frac{\mu_{ab}^*m_jm_k}{(8\pi^2)^2 m_{\Phi^{--}}^2} \, ,
\label{newnm}
\eeq
where $m_j$ denotes the mass of the charged lepton and we assume 
$\Phi^{--}$ is heavier than $\phi^+_a$.
This two-loop neutrino mass matrix is symmetric 
in flavor basis if $\lambda_{--}^{jk}$ is symmetric. This means that only the
symmetric part of $\lambda_{--}^{jk}$ contributes the Majorana neutrino mass.
{}From now on, we assume $\lambda_{--}^{jk}$ is symmetric.

Now, let's estimate the size of $m_\nu^{\rm 2-loop}$ of Eq.~(\ref{newnm}).
\beq
\left(m_\nu^{\rm 2-loop}\right)_{im} \approx 0.5
\sum_{a,b,j,k}\lambda_{--}^{jk}f^a_{ij}f^b_{mk}
\left(\frac{m_jm_k}{m_\tau^2} \right)
\left(\frac{\mu^*_{ab}}{1~{\rm TeV}} \right)
\left(\frac{1~{\rm TeV}}{m_{\Phi^{--}}}\right)^2 \, {\rm keV}.
\eeq
For example, neglecting the electron mass and assuming universal
$\mu_{ab}$,
$\lambda_{--}^{ij}$ and $f^a_{ij}$, the neutrino mass matrix is 
\beq
m_\nu^{\rm 2-loop} \approx 2\,\omega\,\lambda_{--} f^2
\left(
\begin{array}{ccc}
1+2 r_{\mu\tau}+r_{\mu\tau}^2 & 1+r_{\mu\tau}& r_{\mu\tau}+r_{\mu\tau}^2 \\
1+r_{\mu\tau} & 1 & r_{\mu\tau} \\
r_{\mu\tau}+r_{\mu\tau}^2 & r_{\mu\tau} &r_{\mu\tau}^2 
\end{array}
\right)\, {\rm keV},
\eeq
where $r_{\mu\tau}=m_\mu/m_\tau\sim 0.056$, $\lambda_{--}$ and $f$ denote the
universal values of $\lambda^{ij}_{--}$ and $f^a_{ij}$, respectively,
and the normalization factor 
$\omega=
({\mu^*_{ab}}/{1\,{\rm TeV}})\cdot
({1\,{\rm TeV}}/{m_{\Phi^{--}}})^2$.
Thus, it is impossible to explain the large mixing between
$\nu_\mu$ and $\nu_\tau$ with universal $f^a_{ij}$ and
$\lambda^{ij}_{--}$. The hierarchies between $f^a_{ij}$'s and
$\lambda^{ij}_{--}$'s are needed to accommodate the large mixing angle solution
of $\nu_\mu$ for the atmospheric neutrino data.

Since the large (1,1) component of $m_\nu$ is not desirable 
to explain the deficit of $\nu_\mu$ in
Super Kamiokande data as oscillation of $\nu_\mu \rightarrow \nu_\tau$
\cite{zee1},
we need tunings between couplings $\lambda^{ij}_{--}$'s and $f_{ab}$'s. 
For example, we can take $\mu_{ab}$'s as universal parameters. 
To suppress the large (1,1) component of $m_\nu$,
we require the following relations between couplings :
\beqar
\lambda^{23}_{--}&\equiv& \lambda, ~~~~~~~~~
\lambda^{22}_{--}\equiv r_{\mu\tau}^N \lambda , ~~~~~~
\lambda^{33}_{--}= -2 r_{\mu\tau} \lambda , \nonumber \\
f_{12}&\equiv&-f/r_{\mu\tau} , ~~~
f_{13}=-f/r_{\mu\tau} , ~~~
f_{23}\equiv f g \, ,
\label{result}
\eeqar
where we take $N$ as positive integer and neglect the electron mass.
Note that $\lambda^{33}_{--}= -2 r_{\mu\tau} \lambda^{23}_{--}$ and
$f_{12}=f_{13}$.
Then, the mass matrix is given by
\beq
m_\nu^{\rm 2-loop} \approx 2 \,\omega\,\lambda\,f^2 g
\left(
\begin{array}{ccc}
r_{\mu\tau}^N/g    & 1                & 1+r_{\mu\tau}^{N+1} \\
1                  & -2 r_{\mu\tau} g & -r_{\mu\tau} g \\
1+r_{\mu\tau}^{N+1}&-r_{\mu\tau} g    &  r_{\mu\tau}^{N+2} g
\end{array}
\right)\, {\rm keV}.
\eeq
With this mass matrix, we obtain for $N=2$
\beq
\left(
\begin{array}{c}
\nu_e \\ \nu_\mu \\ \nu_\tau
\end{array}
\right) 
\approx
\left(
\begin{array}{ccc}
-\frac{1}{\sqrt{2}}+\frac{g r_{\mu\tau}}{4}  
& \frac{1}{\sqrt{2}}+\frac{g r_{\mu\tau}}{4}
& -\frac{g r_{\mu\tau}}{\sqrt{2}}
\\
\frac{1}{2}+\frac{3\sqrt{2}g r_{\mu\tau}}{8} 
& \frac{1}{2}-\frac{3\sqrt{2}g r_{\mu\tau}}{8} 
& -\frac{1}{\sqrt{2}} 
\\
\frac{1}{2}-\frac{\sqrt{2}g r_{\mu\tau}}{8} 
& \frac{1}{2}+\frac{\sqrt{2}g r_{\mu\tau}}{8} 
& \frac{1}{\sqrt{2}}
\end{array}
\right)
\left(
\begin{array}{c}
\nu_1 \\ \nu_2 \\ \nu_3
\end{array}
\right) \, ,
\eeq
with 
\beqar
m_1^2&\simeq &(2\omega\lambda f^2 g)^2
\left(2+2\sqrt{2}g r_{\mu\tau}\right) ~{\rm keV}^2 ,\nonumber \\
m_2^2&\simeq &(2\omega\lambda f^2 g)^2
\left(2-2\sqrt{2}g r_{\mu\tau}\right) ~{\rm keV}^2 ,\nonumber \\
m_3^2&=&0, 
\eeqar
The mixing matrix and eigenvalues are independent of $N$ if $N > 1$
up to ${\cal O}(r_{\mu\tau})$.
Therefore, to give 
$\Delta m_{\rm atm}^2 = |m_3^2-m_1^2| = |m_3^2-m_2^2| 
\simeq (0.5 - 6) \times 10^{-3} \, {\rm eV}^2$,
the combination of the couplings and mass
parameters satisfies the relation
\beq
 \omega\,\lambda\,f^2 \,g
\simeq (0.8 - 2.7) \times 10^{-5} ,
\eeq
and
\beq
\Delta m_{\rm sol}^2=|m_2^2-m_1^2|=2\sqrt{2}\Delta m_{\rm atm}\, g\, r_{\mu\tau}
\simeq \, (0.8 - 9.5) \times 10^{-5} \, \left(\frac{g}{1/10}\right) \, {\rm eV}^2.
\eeq
We observe that this model can explain the atmospheric neutrino data and
accomodate the large angle MSW resolution of the solar neutrino data.

Now let us consider the phenomenological constraints on $f$'s and $\lambda$'s.
The current experimental data of
the muon decay $\mu \rightarrow \nu_\mu e \bar{\nu}_e$ and 
radiative decay $\mu \rightarrow e \gamma$, whose
constraints $f_{12}$, $f_{13}$, and $f_{23}$, 
are not enough to determine these parameters \cite{zee1,smitani}.
The constraints on $\lambda^{23}_{--}$, $\lambda^{22}_{--}$, and
$\lambda^{33}_{--}$ come from
the tau decay $\tau \rightarrow 3\mu $ and 
radiative decay $\tau \rightarrow \mu \gamma$. 
The decay $\tau \rightarrow 3\mu $ can be described
by the following four-fermion
effective Lagrangian induced by the doubly charged boson exchange (after
appropriate Fiertz transformation) :
\beq
\frac{\lambda_{--}^{ij}\lambda_{--}^{kl*}}{2 m_{\Phi^{--}}^2} \,
\bar{e}_k\gamma^\mu P_R e_i \, \bar{e}_l \gamma_\mu P_R e_j \, .
\eeq
Since the doubly charged scalar interacts only with the right--handed charged
leptons, the effective Lagrangian has $V+A$ form.
{}From this effective Lagrangian we obtain \cite{kkll},
\beq
{\cal B}(\tau \rightarrow 3\mu) \approx
{\cal B}(\tau \rightarrow \nu_\tau e \bar{\nu}_e) 
\left(\frac{\lambda^{33}_{--}\lambda^{22}_{--}}{4\, G_F,
m_{\Phi^{--}}^2}\right)^2.
\eeq
where $G_F$ is the Fermi coupling constant.
{}From the upper bound on the decay mode $\tau \rightarrow 3\mu$ 
and the branching ratio
${\cal B}(\tau \rightarrow \nu_\tau e \bar{\nu}_e)$ \cite{PDG}, we obtain
\beq
\frac{\lambda^{33}_{--}\lambda^{22}_{--}}{m_{\Phi^{--}}^2}
\lesssim 1.3 \times 10^{-2} G_F.
\eeq
This constraints can be easily satisfied without affecting the results from the
neutrino data if, for example, we take large enough $N$ 
even with $\lambda \approx {\cal} O(1)$, viz. Eq. (\ref{result}).
For $\tau \rightarrow \mu \gamma$ through the one--loop diagram with the doubly
charged scalar \cite{zee1},
\beq
{\cal B}(\tau \rightarrow \mu \gamma) \approx
{\cal B}(\tau \rightarrow \nu_\tau e \bar{\nu}_e) \,
\frac{\alpha}{3072\pi}
\left(\frac{\lambda^{33}_{--}\lambda^{23}_{--}}{G_F
m_{\Phi^{--}}^2}\right)^2.
\eeq
{}From the upper bound on the decay mode $\tau \rightarrow \mu \gamma$ \cite{PDG}
and using the Eq. (\ref{result}),
\beq
\frac{r_{\mu\tau}\lambda^2}{m_{\Phi^{--}}^2} \lesssim 2.4 \, G_F.
\eeq
This allows $\lambda \approx {\cal} O(1)$. Therefore, there are no
significant
constraints on $\lambda^{23}_{--}$, $\lambda^{22}_{--}$, and
$\lambda^{33}_{--}$ at present.
But the product of couplings $\omega, \lambda,f$, and $g$
must satisfy the relation~(17). 

\section{Conclusion}

We studied a neutrino mass generating model with a doubly
charged scalar $\Phi^{++}$ and singly charged scalars
$\phi^+_{a,b}$. 
This has been motivated from the observation that
the neutrino mixing angles are large, which is not easily
incorporated in the see-saw mechanism. 
In this study, we emphasized the importance of the consideration of
the lepton number $L$.

However, the parameters introduced in this model
through Eq.~(9) are restricted as Eq.~(17) to 
explain $\Delta m_{\rm atm}^2$.
This constraint is not very
strong since it is a constraint on the product of four coupling
constants.

Generalizing the doubly charged scalar idea to GUT models
such as $SU(5)$ and $SO(10)$ may not be easy. However, it
is not impossible. For example, it can be introduced in higher 
dimensional representations which couple to ${\bf 5}$ of $SU(5)$.
In an $SU(5)\times U(1)$ model, one can introduce a singlet
which is charged. In this case, one does not need high dimensional
representations. However, the GUT inclusion of a doubly
charged scalar field  
is premature to study at present, without a detailed knowledge of
light particles below the GUT scale. 

Distinguishing the doubly charged scalar idea from the see-saw
mechanism can be achieved from the study of rare processes occurring through
the exchange of the doubly charged scalars as studied in this 
paper. Also, if its mass is below the threshold of future accelerators,
it can be easily identified through its decay to four leptons,
$\Phi^{--}\rightarrow l^-l^-\nu\nu$. 

\vspace{1 cm}

While we were finishing this manuscript, we received a preprint by
Joshipura and Rindani \cite{JR} considering  the neutrino masses and mixings
in a Zee model extended by the doubly charged scalar field. In this model,
the neutrino mass matrix of this model has contributions from 
both one-- and two--loop diagrams because the model has one more SM $SU(2)_L$
doublet comparing with the model considered in Sec. III.

\acknowledgments
We thank S. Y. Choi and E. J. Chun for helpful discussions.
This work is also supported in part by KOSEF, MOE through
BSRI 98-2468, and Korea Research Foundation.

\begin{figure}[ht]
\centerline{\epsfig{file=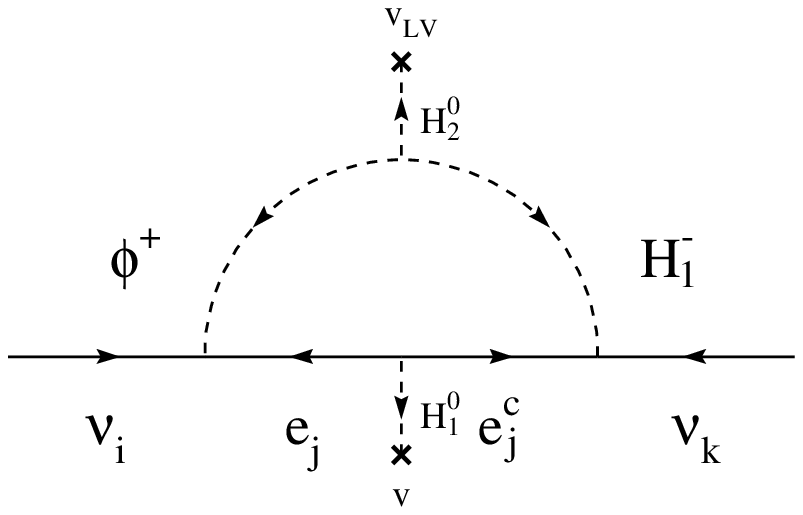,height=12cm, width=20cm}}
\caption{The one-loop neutrino mass in the Zee model.}
\label{olnm_zee}
\end{figure}

\begin{figure}[ht]
\centerline{\epsfig{file=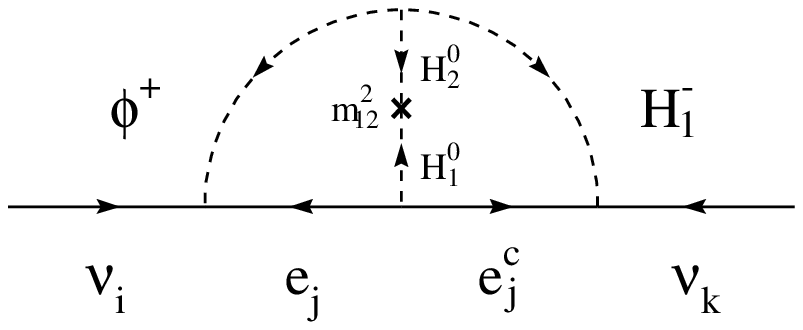,height=12cm, width=20cm}}
\caption{The two-loop neutrino mass in the Zee model.}
\label{tlnm_zee}
\end{figure}

\begin{figure}[ht]
\centerline{\epsfig{file=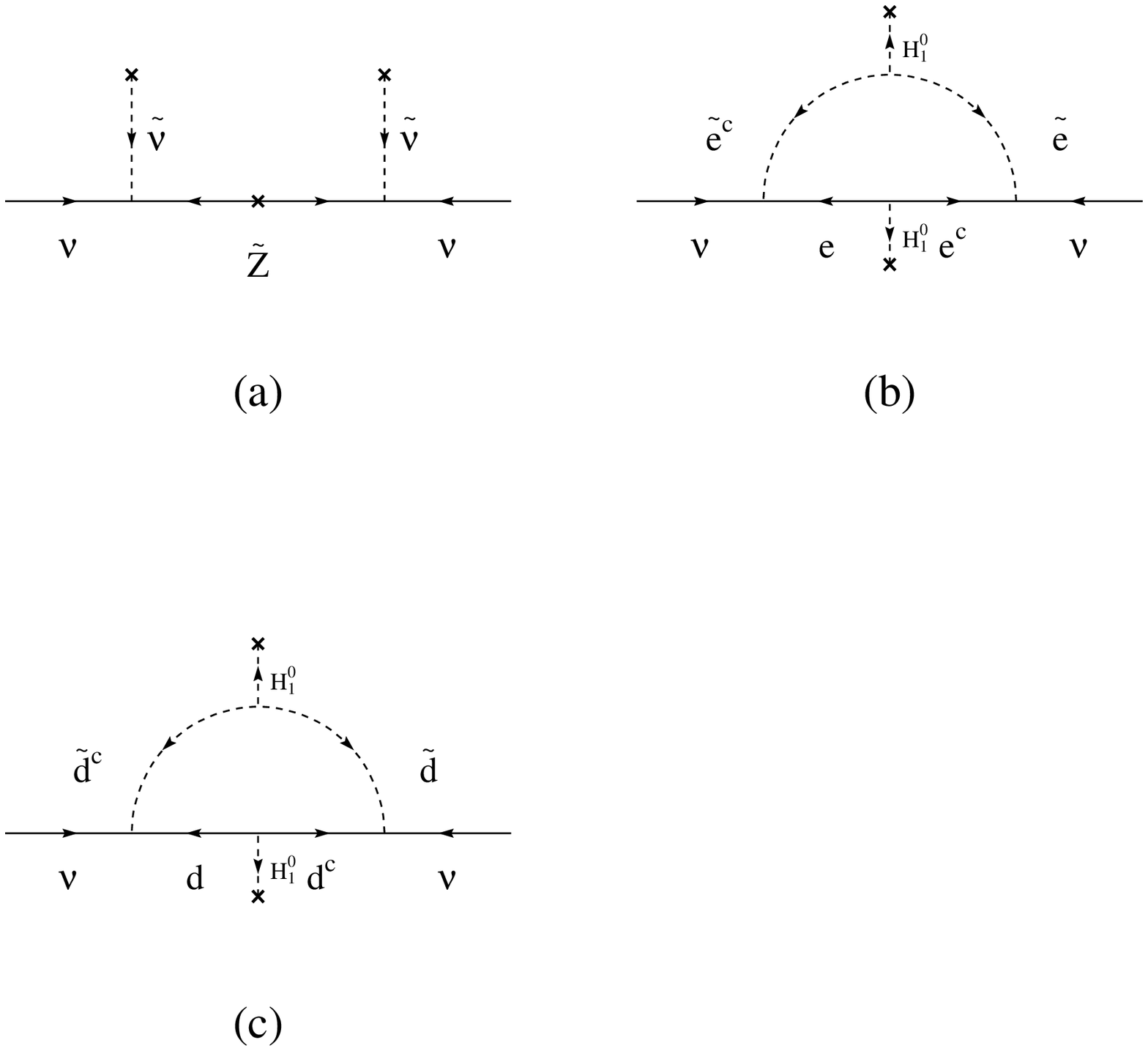,height=20cm, width=20cm}}
\caption{The neutrino mass in the MSSM with R-parity violation.}
\label{rpvnm}
\end{figure}

\begin{figure}[ht]
\centerline{\epsfig{file=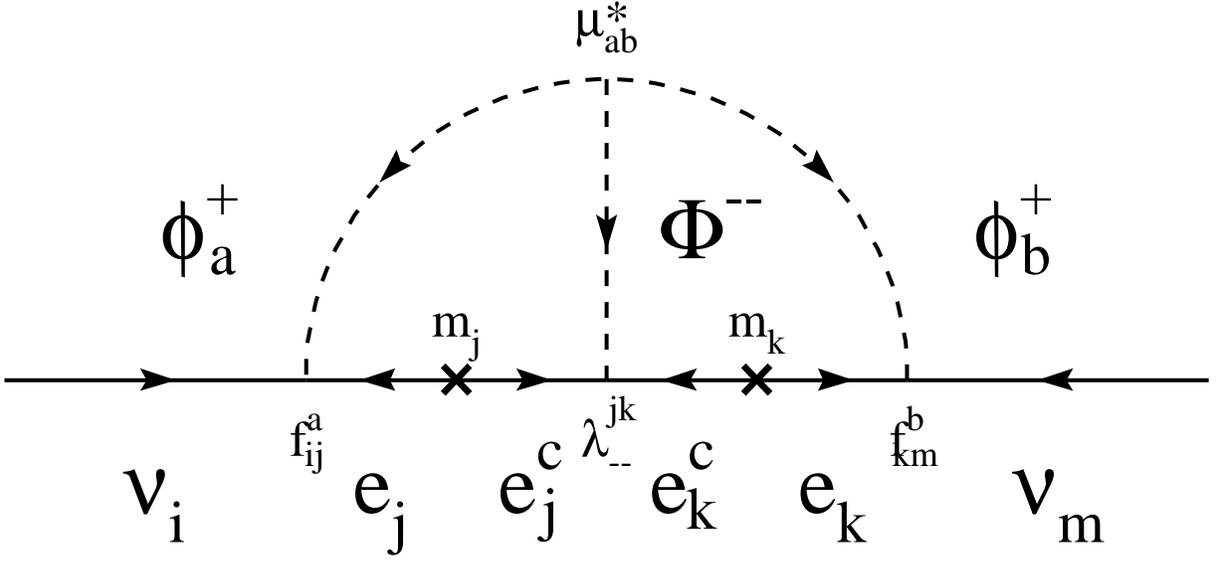,height=12cm, width=20cm}}
\caption{The two-loop neutrino mass in the model with $\phi^+$ and $\Phi^{--}$.}
\label{tlnm}
\end{figure}

\end{document}